\documentclass[sigconf]{acmart}
\usepackage{balance}
\usepackage{bm}
\usepackage{threeparttable}
\usepackage{booktabs} 
\usepackage{multirow}
\usepackage{subfigure}
\usepackage{enumitem}
\usepackage{balance}
\usepackage{enumitem}
\setlist[itemize]{leftmargin=*}
\setcopyright{none}
\usepackage{amsmath} 
 
\usepackage{amssymb}
\usepackage{pdfrender}

\usepackage[hang,flushmargin]{footmisc}

\definecolor{caribbeangreen}{rgb}{0.0, 0.8, 0.6}

\copyrightyear{2021} 
\acmYear{2021} 
\setcopyright{acmcopyright}\acmConference[KDD '21]{Proceedings of the 27th ACM SIGKDD Conference on Knowledge Discovery and Data Mining}{August 14--18, 2021}{Virtual Event, Singapore}
\acmBooktitle{Proceedings of the 27th ACM SIGKDD Conference on Knowledge Discovery and Data Mining (KDD '21), August 14--18, 2021, Virtual Event, Singapore}
\acmPrice{15.00}
\acmDOI{10.1145/3447548.3467384}
\acmISBN{978-1-4503-8332-5/21/08}

\begin{document}
\pagestyle{empty}
\fancyhead{}
\title{Dual Graph enhanced Embedding Neural Network for CTR Prediction}









\author{Wei Guo$^{1\dagger}$, Rong Su$^{1\dagger}$, Renhao Tan$^{1}$$^\ddagger$, Huifeng Guo$^{1}$, Yingxue Zhang$^{2}$, Zhirong Liu$^{1}$, \\ Ruiming Tang$^{1}$*, Xiuqiang He$^{1}$} 
\affiliation{
$^{1}$Huawei Noah's Ark Lab $^{2}$Huawei Noah’s Ark Lab, Montreal Research Center}
\email{{guowei67, surong3, tanrenhao, huifeng.guo, yingxue.zhang, liuzhirong, tangruiming, hexiuqiang1}@huawei.com}


\begin{abstract}
CTR prediction, which aims to estimate the probability that a user will click an item, plays a crucial role in online advertising and recommender system.
Feature interaction modeling based and user interest mining based methods are the two kinds of most popular techniques that have been extensively explored for many years and have made great progress for CTR prediction.
However, (1) feature interaction based methods which rely heavily on the co-occurrence of different features, may suffer from the feature sparsity problem (i.e., many features appear few times); (2) user interest mining based methods which need rich user behaviors to obtain user's diverse interests,  are easy to encounter the behavior sparsity problem (i.e., many users have very short behavior sequences).
To solve these problems, we propose a novel module named \textbf{Dual Graph enhanced Embedding}, which is compatible with various CTR prediction models to alleviate these two problems.
We further propose a \textbf{Dual Graph enhanced Embedding Neural Network} (\textbf{DG-ENN}) for CTR prediction. \textbf{Dual Graph enhanced Embedding} exploits the strengths of graph representation with two carefully designed learning strategies (divide-and-conquer, curriculum-learning-inspired organized learning) to refine the embedding.
We conduct comprehensive experiments on three real-world industrial datasets.
The experimental results show that our proposed \textbf{DG-ENN} significantly outperforms state-of-the-art CTR prediction models. 
Moreover, when applying to state-of-the-art CTR prediction models, \textbf{Dual graph enhanced embedding} always obtains better performance.
Further case studies prove that our proposed \textbf{dual graph enhanced embedding} could alleviate the feature sparsity and behavior sparsity problems.
Our framework will be open-source based on MindSpore\footnote{\noindent{MindSpore. https://www.mindspore.cn/, 2020.}} in the near future.
\noindent\let\thefootnote\relax\footnotetext{$\dagger$~Co-first authors with equal contributions, *~Corresponding author, $^\ddagger$Work done as intern at Huawei Noah's Ark Lab.}
\end{abstract}


\maketitle
{\fontsize{8pt}{8pt}\selectfont
\textbf{ACM Reference Format:}\\
Wei Guo, Rong Su, Renhao Tan, Huifeng Guo, Yingxue Zhang, Zhirong Liu, Ruiming Tang, Xiuqiang He. 2021. Dual Graph enhanced Embedding Neural Network for CTR Prediction. In \textit{Proceedings of the 27th ACM SIGKDD Conference on Knowledge Discovery and Data Mining (KDD’21), August 14–18, 2021, Virtual Event, Singapore}. ACM, New York, NY, USA, 10 pages. https://doi.org/10.1145/3447548.3467384}

\section{Introduction}\label{introduction}
The prediction of click-through rate (CTR) plays a crucial role in many information retrieval (IR) tasks, ranging from web search, personalized recommendation and online advertising, which is multi-billion dollars business nowadays \cite{li2019multi}.
Most of the existing methods for CTR prediction can be classified into two categories, i.e., feature interaction modeling based methods and user interest mining based methods.
Both feature interaction modeling based methods and user interest mining based methods follow a similar Embedding \& Representation learning paradigm: input features are first transformed into trainable embedding vectors which are randomly initialized, and then transformed into fixed-length vector via feature interaction or interest mining, finally fed into fully connected layers to get the prediction score.
Models in the former class such as Factorization Machines (FM) \cite{rendle2010factorization}, Neural FM (NFM) \cite{he2017neural}, Product based Neural Network (PNN) \cite{qu2018product} and FM based Neural Network (DeepFM) \cite{guo2017deepfm} focus on designing novel structure to capture more useful feature interactions more effectively.
The models in the latter class like Deep Interest Network (DIN) \cite{zhou2018deep}, Deep Interest Evolution Network (DIEN) \cite{zhou2019deep} and Multi-Interest Network with Dynamic routing (MIND) \cite{li2019multi} aim to mine user interest from each user's behavior sequence precisely.

\begin{figure}
	\centering
	\setlength{\belowcaptionskip}{-0.3cm}
	\setlength{\abovecaptionskip}{0cm}
	\includegraphics[width=0.45\textwidth]{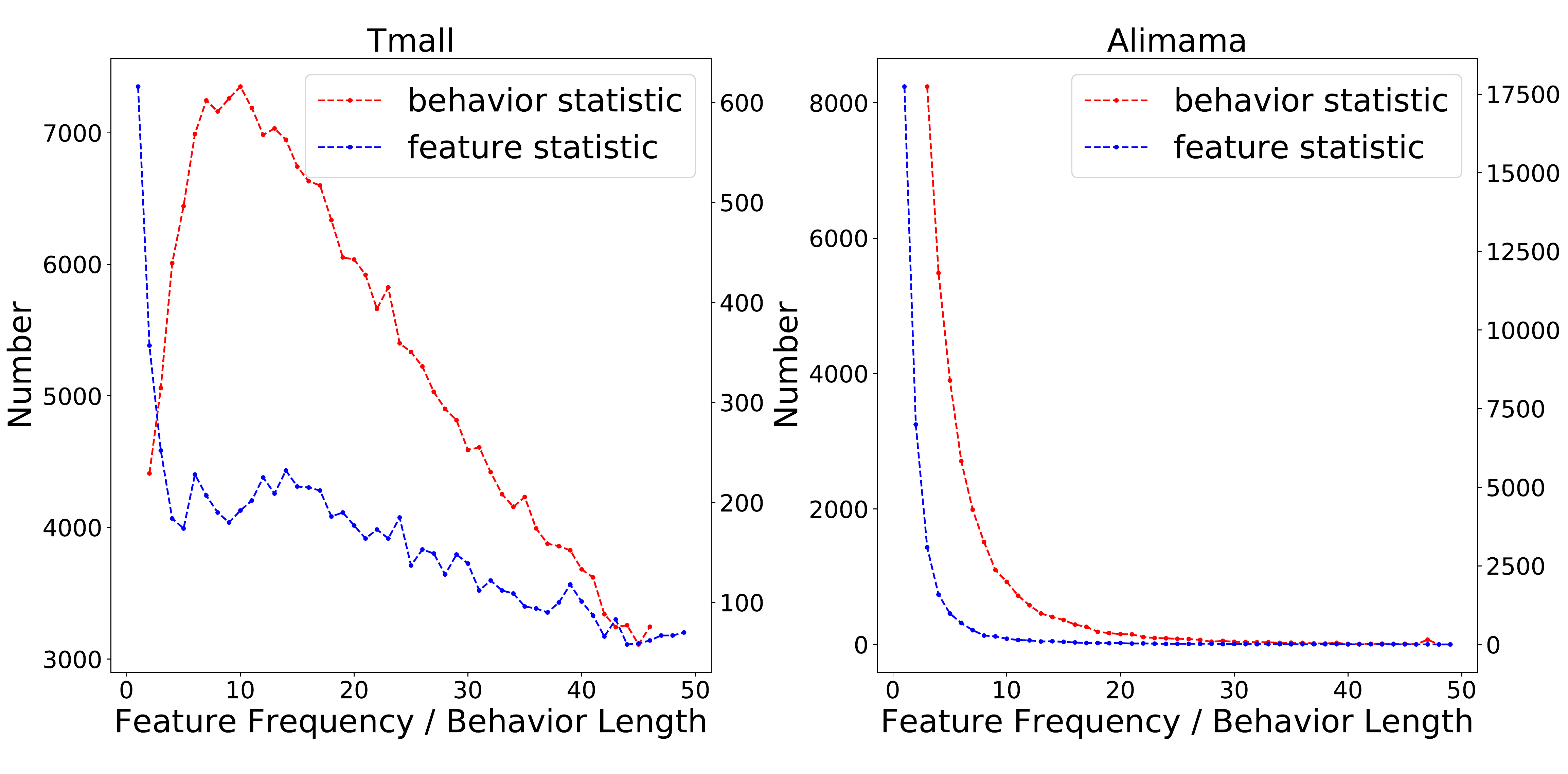}
	\caption{Statistics of feature frequency and behavior length distribution in Tmall and Alimama dataset.
	}
	\label{fig:statistics}
\end{figure}

Though these two kinds of methods for CTR prediction have been investigated for years and obtained great progress, several challenges still exist, which limit the performance of existing methods, especially when deployed in large-scale industrial applications. 
\begin{itemize}
\item \textbf{Feature Sparsity.} The performance of feature interaction based models heavily rely on the co-occurrence of different features.
However, as the number of users and items growing continuously in real recommender system, there are numerous sparse features that appear very few times in the training set.
To verify this discovery, we plot the feature frequency distribution of Tmall\footnote{https://tianchi.aliyun.com/dataset/dataDetail?dataId=42} and Alimama\footnote{https://tianchi.aliyun.com/dataset/dataDetail?dataId=56} dataset in Figure \ref{fig:statistics}.
As we can see, frequency of most features are relatively low.
It is hard for these methods to learn a good representation for these sparse features due to the low frequency of occurrence.


\item \textbf{Behavior Sparsity.} User interest mining based methods need rich user behaviors to obtain user's diverse interests.
However, user behaviors are characterized by heavy-tailed distributions, i.e., a significant proportion of users has very few interactions in the history, as shown in Figure \ref{fig:statistics}. This poses a key challenge  for these models with only limited user behavior information.
\end{itemize}



Recently, graphs have been used to represent relational information in recommendation datasets. Incorporating and exploiting the graph representation has shown to be effective for alleviating data sparsity~\cite{li2019graph,wang2019neural,ying2018graph}. The intuition motivating the involvement of graphs in recommender systems is that we include more relational information and increase the connectivity among users and items, leading to an improvement of recommendation quality.

We propose a novel 
Dual Graph enhanced Embedding Neural Network (DG-ENN), which is designed with two considerations to address the above two challenges in existing methods.
Specifically, we construct two kinds of graphs (i.e., attribute graphs and collaborative graph) from different perspectives to tackle the two above-mentioned sparsity issues. \emph{On the one hand}, a user (item) attribute graph is constructed by using user (item) features, such as gender, age, city, occupation (category, seller, brand). High-order proximity in the attribute graphs helps to enhance the embedding of sparse features, because learning embedding of a node also helps learning embedding of its neighbors, such that sparse features have more chances to be updated. With such enhanced feature embeddings, feature interactions can be learned more effectively. \emph{On the other hand}, a collaborative graph is built from the collaborative signals between users and items. In this graph, there exist edges between a user and an item, representing the user interacting with this item. Besides such user-item edges, user-user edges are defined based on similarity of user profiles and behaviors while item-item edges are formulated based on their transition relations in user behaviors. Exploiting the proximity of this graph, user behaviors with short length can be enhanced with other users' behaviors, by learning node representations.

Yet, how to learn effective user and item representations from the two aforementioned graphs is still challenging due to the following two reasons. First, in the user (item) attribute graph, the user (item) attributes are of different fields with very different characteristics, which makes aggregating them during the learning process non-trivial. Second, in the collaborative graph, there are various kinds of edges, resulting complex relations such as $u_1 \rightarrow v_1$, $u_1 \rightarrow u_2 \rightarrow v_1$, $u_1 \rightarrow u_2 \rightarrow v_2 \rightarrow v_1$ and $u_1 \rightarrow v_2 \rightarrow u_2 \rightarrow v_1$, which makes the relation modeling between user $u_1$ and item $i_1$ difficult.
To handle such two issues, DG-ENN learns the user and item embeddings from these two graphs with two novel strategies. To learn embeddings in the user (item) attribute graph, a divide-and-conquer strategy is proposed to learn the information for each field of attributes individually and perform the information integration at the end, so that the information of different attributes (with different semantics) will not make the learning process chaotic. When learning from the collaborative graph, an organized learning mechanism, inspired by curriculum learning, is introduced to learn the user-user and item-item edges (which are relative easier to train) first, and learn user-item edges after that.
Furthermore, DG-ENN serves as an embedding learning framework, which works compatibly with the existing deep CTR models, including both feature interaction modeling based and user interest mining based methods.
To sum up, our contributions in this paper can be summarized as follows: 
\begin{itemize}
	\item We propose a novel Dual Graph enhanced Embedding Neural Network named DG-ENN, which enhances the feature embedding in an end-to-end graph neural network framework. To the best of our knowledge, this is the first deep CTR model using graphs for alleviating the feature sparsity and behavior sparsity problems. 
	


	\item More specifically, a user (item) attribute graph and a collaborative graph in DG-ENN are proposed to alleviate the feature sparsity and behavior sparsity problem. To learn these graphs effectively, we propose to perform a divide-and-conquer learning strategy and a curriculum-learning-inspired organized learning strategy for these two kinds of graphs, respectively.
	

	
	\item We perform extensive experiments on three public datasets, demonstrating significant improvements of DG-ENN over state-of-the-art methods for CTR prediction. The necessity of the two kinds of graphs is verified empirically. Moreover, the validity of the proposed two learning strategies is also demonstrated.
\end{itemize}

\section{Related Work}\label{related_work}
We briefly review three kinds of existing methods that are relevant to our work: 1) feature interaction modeling for CTR prediction, 2) user interest mining for CTR prediction, and 3) graph neural network for recommendation.
\newline
\textbf{Feature Interaction Modeling for CTR prediction.}
Using raw features directly for CTR prediction can hardly lead to a good result, thus feature interactions modeling is playing a core role and has been extensively studied in the literature \cite{lian2018xdeepfm}.
FM utilizes a low dimensional latent vector to represent each feature and learns 2-order feature interactions through the inner product of the related features' vectors \cite{rendle2010factorization}.
Owing to its superior performance in learning feature interactions, many extensions of FM are proposed \cite{juan2016field,pan2018field, xiao2017attentional}.
Recently, Deep Neural Network (DNN) has achieved great success with its great power of feature representation learning.
It is promising to exploit DNN for CTR prediction.
NFM \cite{he2017neural} enhances FM with DNN to model non-linear and high-order feature interactions simultaneously.
PNN further introduces a product layer between the embedding layer and DNN to model the feature interactions \cite{qu2018product}.
Wide \& Deep \cite{cheng2016wide} and DeepFM \cite{guo2017deepfm} introduce an interesting hybrid architecture, which contain a shallow model and a DNN model to learn low-order and high-order feature interactions simultaneously.
Deep \& Cross Network (DCN) \cite{wang2017deep} and  CIN \cite{lian2018xdeepfm} apply feature crossing at each layer explicitly. Thus the orders increase at each layer and are determined by layer depth.
\newline
\textbf{User Interest Extraction for CTR prediction.}
Besides feature interactions modeling, user interest extraction is also very important.
Many works are proposed recently that focus on learning user interest representation from user behavior history.
DIN supposes that user interest is diverse, then uses an attention network to assign different scores to different user behaviors for user representation learning \cite{zhou2018deep}.
DIEN observes that user interest is dynamic, thus it utilizes GRU layers and auxiliary loss to capture evolving user interest for user's historical behavior sequence \cite{zhou2019deep}.
DSIN argues that user behavior sequence are composed of different homogeneous sessions \cite{feng2019deep}.
So it employs self-attention layer and Bi-LSTM to model user's inter-session and intra-session interests.
MIND learns multiple vectors for representing user's interests by using capsule network and dynamic routing architecture \cite{li2019multi}.
Despite great success has been made by these two kinds of CTR prediction methods, they cannot effectively solve the feature sparsity and behavior sparsity problems.
We are going to solve them in this paper by incorporating and exploiting the graph representation learning.
\newline
\textbf{Graph Neural Network for Recommendation.}
Graph Neural Network is widely used in recommender system in recent years.
FiGNN models feature interactions via graph propagation on the fully-connected fields graph \cite{li2019fi}.
GIN utilizes user behaviors to construct a co-occurrence commodity graph to mine user intention \cite{li2019graph}.
GCMC \cite{berg2017graph} treats the recommendation task as a link prediction problem and employs a graph auto-encoder framework on the user-item bipartite graph to learn user and item embeddings.
To better capture the collaborative signal existed in the user-item bipartite graph, many other GNN based works are then be proposed \cite{ying2018graph, wang2019neural, he2020lightgcn}.
To make full use of other information beyond user-item interactions, KGAT \cite{wang2019kgat} constructs a collaborative knowledge by combining user-item graph with knowledge graph and then applies graph convolution to get the final node representations. 
Heterogeneous graph Attention Network (HGAT) \cite{linmei2019heterogeneous} utilizes a semantic-level attention network and a node-level attention network to discriminate the importance of neighbor nodes and node types. 
Although these GNN-based models have made progress, applying them directly for CTR prediction is still challenging, as depicted in Section \ref{introduction}.

\section{Preliminary}
\subsection{Problem Definition}
In this section, we formulate the CTR prediction task with necessary notations.
There are a set of $M$ users $\mathcal{U} = \left\{u_1,u_2,...,u_M\right\}$, a set of $N$ items $\mathcal{V} = \left\{v_1,v_2,...,v_N\right\}$, a set of $J$ fields of user attributes $\mathcal{A} =  \left\{\mathcal{A}^1,\mathcal{A}^2,...,\mathcal{A}^J\right\}$, a set of $K$ fields of item attributes $\mathcal{B} =  \left\{\mathcal{B}^1,\mathcal{B}^2,...,\mathcal{B}^K\right\}$ and a set of $R$ fields of other features like timestamp, displayed position denoted as $\mathcal{C} =\left\{\mathcal{C}^1,\mathcal{C}^2,\cdots,\mathcal{C}^R\right\}$ to describe the context. 
The user-item interactions are denoted as a matrix $\mathcal{Y}\in\mathbb{R}^{M \times N}$, where $y_{uv} = 1$ denotes user $u$ has interaction with item $v$ before, otherwise $y_{uv}=0$.
Further, each user and item is associated with a list of attributes $\mathbf{A}_u \subset \mathcal{A}$ and $\mathbf{B}_v \subset \mathcal{B}$.
In addition to the attributes, each user also has a behavior sequence denoted as $\textbf{S}_u = \{s_1,s_i,\cdots,s_{T_u}\}$, where $s_i \in \mathcal{V}$ and $T_u$ is the length of user $u$'s behavior sequence in the past.
Besides user and item features, we denote context features as a list of $\mathbf{C} \subset \mathcal{C}$.
Concatenating all these features in a predefined order, one instance can be represented as: 
\begin{equation}
\textbf{x}= [u, v, \textbf{A}_u, \textbf{B}_v, \textbf{S}_u, \textbf{C}]
\end{equation}
An encoding example of user ID, user attribute and user behavior feature is presented as:
$$
 \begin{matrix}
 \underbrace{ [0,0,\cdots,1,0] } &  \underbrace{ [0,1,\cdots,1] } &  \underbrace{ [1,1,\cdots,1,0] } \\
 u: User ID & \textbf{A}_u:Occupation \& City &  \textbf{S}_u: Click
\end{matrix}
$$
The representations of other features are similar, so we omit them for simplicity.
It is noticed that we categorize user id, item id, user attributes, item attributes and context as features except the user behavior.
These features may encounter the feature sparsity problem when appear very few times.  
The goal of CTR prediction is to predict the probability that user $u$ will be interested in the target item $v$ under context $\textbf{C}$.



\subsection{Base model}

Most of existing CTR prediction methods follow a similar Embedding \& Representation learning \& Prediction paradigm.
We refer them as the base model in this section.

\subsubsection{Initial Embedding}
The input data in CTR prediction are usually in a high-dimensional sparse binary form. 
It is common to apply an embedding layer upon the input to compress it into a low dimensional, dense real-value vector by looking up from an embedding table \textbf{W}.
For one-hot vector $u,v$, the embedding representation is a single vector.
For multi-hot vector $\textbf{A}_u, \textbf{B}_v, \textbf{S}_u, \textbf{C}$, the embedding representation is a list of vectors.
The embedding vectors of these fields are then concatenated together to get the embedding of the whole input features.
\begin{equation}
 \textbf{E} = [\textbf{e}_u, \textbf{e}_v, \textbf{e}_{\textbf{A}_u}, \textbf{e}_{\textbf{B}_v}, \textbf{e}_{\textbf{S}_u}, \textbf{e}_{\textbf{C}}]
 \label{output_emb}
 \end{equation}
 
\subsubsection{Representation Learning}
Many existing works focus on designing advanced network architecture for feature interaction modeling or user interest mining, which can be formulated as:
\begin{align}
\textbf{P} = f(\textbf{E})
\end{align}
For simplicity, we use the inner product module as the base representation learning module:
\begin{align}
\textbf{P} = [\textbf{E}_1, \cdots, \textbf{E}_F, \langle\textbf{E}_1,\textbf{E}_2\rangle,\cdots, \langle\textbf{E}_{F-1},\textbf{E}_{F}\rangle]
\end{align}
where $\langle,\rangle$ denotes the inner product operation, $F$ is the number of fields.
We also evaluate the performance of other representation learning module in the experiment section to validate the effectiveness of our proposed dual graph enhanced embedding.

\subsubsection{Fully Connected Layer}
The output of representation learning component is fed into the fully connected layer, which serves as a classifier.
\begin{equation}
 \textbf{a}^{(l)} = \sigma(\textbf{W}^{(l)}\textbf{a}^{(l-1)}+\textbf{b}^{(l)})
 \end{equation}
 where $\textbf{a}^{(0)} = \textbf{P}$, $l$ is the current layer depth and $\sigma$ is the activation function. $\textbf{a}^{(l-1)}$, $\textbf{W}^{(l)}$ and $\textbf{b}^{(l)}$ are the input, model parameters and bias of the $l$-th layer. The output is a real number as the predicted CTR:
\begin{equation}
\widehat{y} = Sigmoid(\textbf{W}^{(L)}\textbf{a}^{(L-1)}+\textbf{b}^{(L)})
 \end{equation}
 
\subsubsection{Model Training}
The widely-used logloss is adopted as the objective function, which is defined as:
\begin{equation}
L_{logloss} = -\frac{1}{|\mathcal{S}|}\sum_{(\textbf{x}, y) \in \mathcal{S} }{y(\textbf{x})log\widehat{y}(\textbf{x})+(1-y(\textbf{x}))log(1-\widehat{y}(\textbf{x}))}
 \end{equation}
where $|\mathcal{S}|$  is the total number of training instances, $y(\textbf{x})$ is the real value for input vector $\textbf{x}$, and $\widehat{y}(\textbf{x})$ is the predicted value by our model.


\section{Dual Graph enhanced Embedding}

\begin{figure*}
	\centering
	\setlength{\belowcaptionskip}{-0.3cm}
	\setlength{\abovecaptionskip}{0cm}
	\includegraphics[width=1\textwidth]{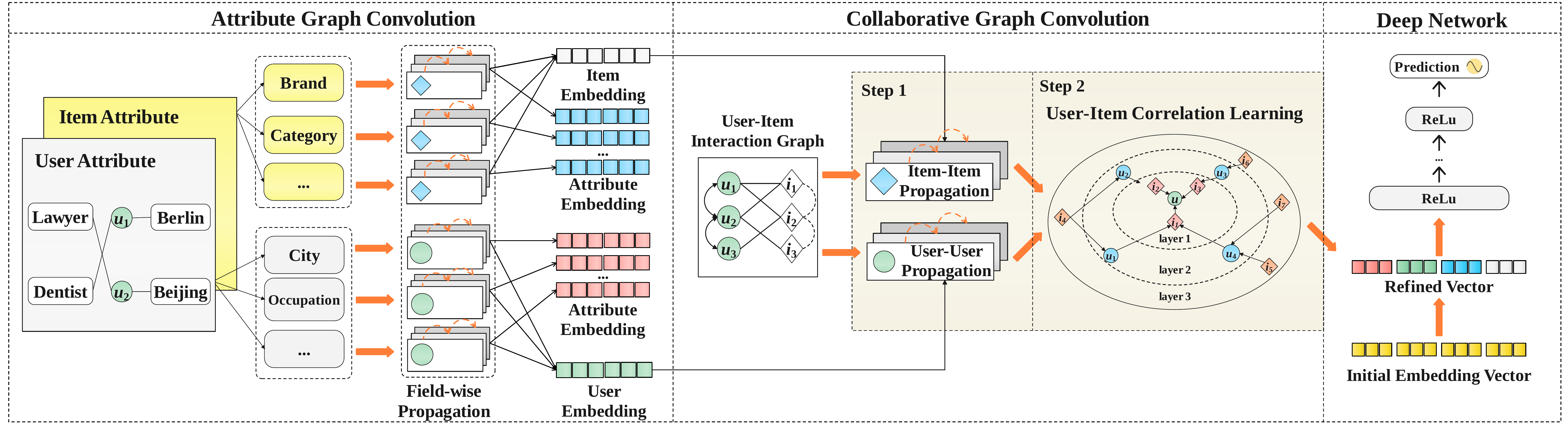}
	\caption{Overview of our proposed DG-ENN framework. The left part is the attribute graph convolution module, the central part is the collaborative graph convolution module and the right part is the deep network module.} 
	\label{fig:model}
\end{figure*}
As stated earlier, most of existing methods focus on the representation learning layer, while overlook the embedding layer. Whereas, embdding layer with random initialization suffers from the feature sparsity and behavior sparsity issue.
Motivated by this observation, in this paper, we focus on the embedding learning  with a dual graph enhanced embedding network (DG-ENN) based on the base model.

Dual graph enhance embedding component contains three modules: \textbf{graph construction}, \textbf{attribute graph convolution} and \textbf{collaborative graph convolution}. In this section, we elaborate each of these three modules in detail. Figure \ref{fig:model} gives a depiction of our proposed dual graph convolution framework.
\subsection{Graph Construction}
\subsubsection{Attribute Graph} 
An attribute can be in multiple users or items, serving as a bridge to improve their representation.
Based on this bridge, we construct two attribute graphs $\mathcal{G}_{ua}=\langle \mathcal{U}\cup\mathcal{A},  \mathcal{E}_{ua}\rangle$ and $\mathcal{G}_{vb}=\langle \mathcal{V}\cup\mathcal{B}, \mathcal{E}_{vb}\rangle$.
Edges $e_{ua}=(u,a)$ and $e_{vb}=(v,b)$ indicate that attribute $a$ belongs to user $u$ and attribute $b$ belongs to item $v$.
The attribution graphs establish attribute connections to alleviate the feature sparsity problem. 
\subsubsection{Collaborative Graph.}

Inspired by the collaborative filtering (CF) that similar users may exhibit similar preference on items \cite{sarwar2001item}, we utilize the collaborative signals to expand user behaviors and therefore alleviate the behavior sparsity problem. 
User-item interactions matrix $\mathcal{Y}$ can be regarded as a user-item bipartite graph $\mathcal{G}_{uv}=\langle \mathcal{U}\cup\mathcal{V}, \mathcal{E}_{uv}\rangle$. 
There is an edge $e_{uv}=(u,v)$ if $y_{uv} = 1$. 
However, $\mathcal{G}_{uv}$ only reveals the user-item interaction relation, but ignores the direct connections inside users and inside items. 
As a result, we construct user-user similarity graph and item-item transition graph to extract such more complex relations.
The user-user similarity graph is built based on the user preferences and user attributes simultaneously:
\begin{align}
    sim(i,j) = \alpha_1\frac{\langle\textbf{Y}_i,\textbf{Y}_j\rangle}{||\textbf{Y}_i|| \cdot ||\textbf{Y}_j||} + \alpha_2\frac{\langle\textbf{A}_i,\textbf{A}_j\rangle}{||\textbf{A}_i|| \cdot ||\textbf{A}_j||}
\end{align}
where $\textbf{Y}_i$ and $\textbf{Y}_j$ denote the $i$-th and $j$-th row of the user-item interaction matrix $\mathcal{Y}$, $\textbf{A}_i$ and $\textbf{A}_j$ denote the attributes of corresponding user $i$ and user $j$.
We set $\alpha_1 = \alpha_2 = 0.5$ for simplicity. 
After calculating the overall similarity between each two users, we can build a $k$-NN graph $\mathcal{G}_{uu}=\langle \mathcal{U}\cup\mathcal{U}, \mathcal{E}_{uu}\rangle$ with a pre-defined $k$. 
The item-item transition graph is built based on the sequential information of different users' behavior sequences.
Two items are connected in the item-item transition graph if they are interacted by the same users consecutively.
With all users' behavior sequences considered, we can construct an item-item graph $\mathcal{G}_{vv}=\langle \mathcal{V}\cup\mathcal{V}, \mathcal{E}_{vv}\rangle$. It can reflect user's preferences on group of items, which are ignored by the user-item bipartite graph. 
As a result, we can get the overall user-item collaborative graph $\mathcal{G}_{cf}=\langle \mathcal{U}\cup\mathcal{V}, \mathcal{E}_{uu}\cup\mathcal{E}_{uv}\cup\mathcal{E}_{vv}\rangle$.
By aggregating neighborhood information from $\mathcal{G}_{cf}$ iteratively, user's representation can be enhanced with other users’ behaviors, thus the behavior sparsity problem can be alleviated. 



\subsection{Attribute Graph Convolution}\label{Attribute Graph Convolution}
With the two attribute graphs $\mathcal{G}_{ua}$ and $\mathcal{G}_{vb}$, we enrich the representation of sparse features with graph convolution.
The user (item) attributes contain different fields with very different characteristics (for example, item price and item category are very different in semantics as well as distributions), which makes aggregating them during the learning process non-trivial.

However, most of existing GNNs mix neighbors information indistinguishably and fail to distinguish different characteristics of neighbor attributes nodes, leading to sub-optimal results \cite{kipf2016semi,wang2019neural,he2020lightgcn}.
To consider different characteristics of attributes, we propose a divide-and-conquer strategy to integrate different attribute information while maintaining their intrinsic characteristics.
More specifically, we learn the information for each field of attributes individually and perform information integration at the end.
\subsubsection{Field-wise Information Propagation}\label{Field-wise Information Propagation}
We first describe the information propagation within a field of attributes.
We adopt the state-of-the-art GCN models for such field-wise information propagation.
We use $h$ to denote the central node and $N_h$ to denote its neighbor set in this graph.
We adopt the following three types of GCN aggregators as potential candidates:
\begin{itemize}
\item \textsl{GCN Aggregator.}
GCN \cite{kipf2016semi} sums up the representation of central node and its neighbors and then applies a nonlinear transformation to generate the new representation:
\begin{align}
f_{GCN}^{(l)}(\textbf{e}_h^{(l)}, \textbf{e}_{N_h}^{(l)}) = \sigma (W^{(l)} \sum_{i\in \{h\} \cup \{N_h\}} d(h,i) \textbf{e}_i^{(l)})
\label{GCN}
\end{align}
where $\mathbf{e}^{(0)}$ is the initial embedding from $\mathbf{E}$, 
$\sigma$ and $W^{(l)}$ are the nonlinear activation function and transformation matrix of layer $l$.
$d(h,i) = 1/\sqrt{|N_h||N_i|}$ is the normalization factor. 
\item \textsl{NGCF Aggregator.} NGCF \cite{wang2019neural} improves GCN by considering additional feature interactions between central node and neighbor nodes. Besides, it aggregates the neighbors first and then add the neighbor representation to central representation, which can be formulated as follows:
\begin{equation}
\begin{aligned}
f_{NGCF}^{(l)}(\textbf{e}_h^{(l)}, \textbf{e}_{N_h}^{(l)}) = \sigma(W_1^{(l)}\mathbf{e}_{h}^{(l)} + \sum_{i\in N_h} d(h,i)(W_1^{(l)}\textbf{e}_i^{(l)} + \\ W_2^l(\textbf{e}_h^{(l)} \odot \textbf{e}_{i}^{(l)})))
\label{NGCF}
\end{aligned}
\end{equation}
where $W_1^{(l)}$, $W_2^{(l)}$ are the trainable weight matrix and $\odot$ denotes the element-wise product. 
\item \textsl{LightGCN Aggregator.} LightGCN \cite{he2020lightgcn} argues that the feature transformations and nonlinear activation function are not necessary and might even degrade the recommendation performance. Therefore, it removes the weight matrix and activation function:
\begin{align}
f_{LightGCN}^{(l)}(\textbf{e}_h^{(l)}, \textbf{e}_{N_h}^{(l)})  = \sum_{i\in N_h} d(h,i) \textbf{e}_i^{(l)} 
\label{lightgcn}
\end{align}
\end{itemize}

The feature representation with layer $l+1$ information propagation is formulated as:
\begin{align}
\textbf{e}_h^{(l+1)} = f_{\star}^{(l)}(\textbf{e}_h^{(l)}, \textbf{e}_{N_h}^{(l)})
\end{align}
Noticed that we use separate parameters for different fields when using GCN or NGCF aggregators. As aggregators are very important for the performance of our method, we will evaluate the effectiveness of the three GCN aggregators in experiment section.
After propagation of $L$ layers, we have $L + 1$ embeddings for each node.
Following \cite{he2020lightgcn}, we average these embeddings to get the final embedding for all central nodes:
\begin{align}
\hat{\textbf{e}}_h = \sum_{l=0}^{L}\textbf{e}_h^{(l)}
\end{align}

\subsubsection{Cross-field Information Integration.}

As we have $J$ fields of user attributes and $K$ fields of item attributes, we generate $J$ user representations and $K$ item representations by field-wise information propagation in the previous section.
As different fields of attributes have different importance to the final representation, it's natural to employ an attention mechanism to assign different importance scores for individual representations.
However, as the main contribution of this part of model is introducing the attribute graphs and modeling field-wise information individually (the effectiveness of which will be validated empirically in the experiments), we apply the average operation over multiple embeddings to get the final user and item representations:
\begin{align}
\textbf{z}_u = \sum_{t=1}^{J}\hat{\textbf{e}}_u^{(t)},  
\textbf{z}_v = \sum_{t=1}^{K}\hat{\textbf{e}}_v^{(t)}
\end{align}
where $\hat{\textbf{e}}_u^{(t)}$ denotes the embedding obtained from Section \ref{Field-wise Information Propagation} with respect to user attribute of field $t$ (we omit $t$ in Section \ref{Field-wise Information Propagation} for the sake of clarity).
The embedding of all the features in data instance can be refined as:
\begin{equation}
 \textbf{Z} = [\textbf{z}_u, \textbf{z}_v, \textbf{z}_{\textbf{A}_u}, \textbf{z}_{\textbf{B}_v}, \textbf{z}_{\textbf{S}_u}, \textbf{e}_{\textbf{C}}]
 \end{equation}
Noted that all the features (except contextual features) are enhanced.
The reason why we don't construct graphs for contextual features is the risk of introducing noise as there are no clear relations between users/items and contextual features in most cases.



\subsection{Collaborative Graph Convolution}\label{Collaborative Graph Convolution}
The behavior sparsity issue is a challenge for the model to capture user interests with very limited user behavior information.
Using the high-order proximity of the collaborative graph to enrich user behaviors is beneficial to alleviate this issue.
However, the underlying reasons motivating a user to click an item may be various, which might be difficult for existing models to capture such complex relations.
For example, $u_1 \rightarrow v_1$, $u_1 \rightarrow u_2 \rightarrow v_1$, $u_1 \rightarrow u_2 \rightarrow v_2 \rightarrow v_1$ and $u_1 \rightarrow v_2 \rightarrow u_2 \rightarrow v_1$
are all possible reasons to drive user $u_1$ to click on a target item $v_1$.
Existing meta-path based methods, like \cite{hu2018leveraging,wang2019heterogeneous}, introduce additional information with the path extraction strategy.
However, they need expert knowledge to design meta-paths.
Besides, it's difficult for meta-path methods to exhaustively search all useful meta-paths, which largely limits their performance.
GNN based methods use neighbor aggregation for behavior expanding, which don't need domain knowledge.
However, existing GNN based methods \cite{wang2019kgat,linmei2019heterogeneous,zhang2019heterogeneous} aggregate different types of neighbors at the same time, which overlook the dependency during the process of neighbor aggregation.
This reduce their ability to model graph correlations and more complex graph structure.
To solve these issues, we design an organized learning mechanism by taking inspiration from the curriculum learning, which introduces different concepts at different time and then uses the previous learned concepts to promote the learning of new concepts \cite{bengio2009curriculum}.
Concretely, we first learn separate representations for users and items using user-user edges and item-item edges, then we use the user-item edges to learn the correlations between users and items. 
By this way, the complex node relation can be modeled well.
As shown in the right part of Figure \ref{fig:model}, collaborative graph convolution network includes two components: 1) information propagation within Users/Items, 2) behavior expanding across users and items. 


\subsubsection{Information Propagation within Users/Items}
We first illustrate the information propagation within users/items. The input of this component is the refined embedding from attribute graph convolution module.
Taking user node as an example, we denote the central node as $u$ and its user neighbor set as $N_u$. 
The information propagation is formulated as: 
\begin{align}
\textbf{z}_u^{(l)} = f_{\star}^{(l-1)}(\textbf{z}_u^{(l-1)}, \textbf{z}_{N_u}^{(l-1)})
\end{align}
where $\textbf{z}_0^{(0)}$ and $\textbf{z}_{N_u}^{(0)}$ are the refined embedding from $Z$.
We use the average pooling of all layer's output as the final representation, as different layers of information propagation can represent different length of relations:
\begin{align}
\hat{\textbf{z}}_u = \sum_{l=0}^{L}\textbf{z}_u^{(l)}
\end{align}
Similarly, the representation for item nodes is:
\begin{align}
\hat{\textbf{z}}_v = \sum_{l=0}^{L}\textbf{z}_v^{(l)}
\end{align}


\subsubsection{Behavior Expanding Across Users and Items}
After learning from user-user and item-item edges, 
we use user-item edges to learn the user-item preferences that can be used for user behavior expanding.
Taking user node as an example, the user-item correlations can be modeled as: 
\begin{align}
\textbf{q}_u^{(l)} = f_{\star}^{(l-1)}(\textbf{q}_u^{(l-1)}, \textbf{q}_{N_u}^{(l-1)})
\end{align}
where $\textbf{q}_u^{(0)} = \hat{\textbf{z}}_u$ and $\textbf{q}_{N_u}^{(0)} = \hat{\textbf{z}}_{N_u}$ are the enriched embedding after information propagation within users/items.
Then we use the average pooling of all layers' output as the final representation:
\begin{align}
\hat{\textbf{p}}_u = \sum_{l=0}^{L}\textbf{p}_u^{(l)}
\end{align}
Similarly, the embedding of item $v$ is generated by the same process:
\begin{align}
\hat{\textbf{p}}_v = \sum_{l=0}^{L}\textbf{p}_v^{(l)}
\end{align}
Notice that $\textbf{q}_u^{(0)} = \hat{\textbf{z}}_u$ and $\textbf{q}_v^{(0)} = \hat{\textbf{z}}_v$ are also included in the final representation, because user-user relations and item-item relations also contain useful neighbors that can be used to expanding user behaviors.
Comparing with equation~\ref{output_emb}, after the graph enhanced operations,  we can get the final enhanced embdding for all the features:
\begin{equation}
 \textbf{P} = [\hat{\textbf{p}}_u, \hat{\textbf{p}}_v, \textbf{z}_{\textbf{A}_u}, \textbf{z}_{\textbf{B}_v}, \hat{\textbf{p}}_{\textbf{S}_u},
 \textbf{e}_{\textbf{C}}]
 \end{equation}

\subsection{Complexity Analysis}
Since scalability is important for graph-based algorithms, we analyze the time complexity of DG-ENN for model training and online inference respectively.
As the enhanced embedding can be used directly for online inference, the time complexity of DG-ENN is \textbf{the same as} base model.
For model training, the layer-wise graph convolution is the main time cost.
Taking LightGCN aggregator as an example, the computational complexity for attribute graph is $\mathcal{O}(L \cdot |\mathcal{G}_{ua}| \cdot d$), where $|\mathcal{G}_{ua}|$ denotes the number of edges existed in $\mathcal{G}_{ua}$, $d$ is the embedding size and $L$ is the number of graph convolution layers.
Similarly, the computational complexity for collaborative graph is $\mathcal{O}(L \cdot |\mathcal{G}_{cf}| \cdot d$).
In real-world industrial application, there may be numerous edges connecting users (items) with attributes and connecting users and items.
To scale up the model training, neighbor sampling is necessary.

\section{EXPERIMENTS}\label{experiment}
\subsection{Experiment Setup}\label{ExperimentSetup}
\subsubsection{Datasets}
We evaluate the effectiveness of our proposed model on three large-scale datasets: \emph{Alipay}, \emph{Tmall}, and \emph{Alimama}.
\begin{itemize}
	\item \textbf{Alipay}\footnote{https://tianchi.aliyun.com/dataset/dataDetail?dataId=53}: 
	This dataset is provided by Ant Financial Services in IJCAI-16 contest \cite{qin2020user}.
	It contains users' online/on-site behavior logs in 2015.
	Each log contains multiple fields, including user ID, item ID, seller, category, online action type and timestamp.
	\item \textbf{Tmall}\footnote{https://tianchi.aliyun.com/dataset/dataDetail?dataId=42}: 
	This dataset is provided by Tmall.com in IJCAI-15 contest \cite{qin2020user}.
	The user profile is described by user ID, age range and gender. The item attributes include category and brand. The context features are timestamp and action type.
	\item \textbf{Alimama}\footnote{https://tianchi.aliyun.com/dataset/dataDetail?dataId=56}: 
	This dataset is provided by Alimama \cite{feng2019deep}.
	Each log in this dataset is composed of 12 feature fields including user ID, item ID, user micro group ID, occupation, shopping level, brand, category and some other information.
\end{itemize}
\subsubsection{Dataset Preprocessing.}
For each user, their clicked items are sorted by the interaction timestamp.
Following \cite{ren2019lifelong,qin2020user}, we split the dataset for evaluation.
Specifically, supposing there are \textbf{T} historical behaviors for a user, behavior [\textsl{1}, \textsl{T-3}] are collected as user behavior feature in the training set to predict the target item \textsl{T-2}. 
Similarly, behavior [\textsl{1}, \textsl{T-2}] are used as user behavior feature in the validation set to predict the target item \textsl{T-1}, behavior [\textsl{1}, \textsl{T-1}] are used as user behavior feature in the testing set to predict the target item \textsl{T}.
For each user, we random sample 10 non-clicked items to replace the target item as the negative samples.
Table \ref{tab:dataset} shows the statistics of the three datasets. 
 
\begin{table}[t]
 \caption{\small{Dataset statistics.}}
 \centering
 	\vspace{-0.2cm}
 	\setlength{\tabcolsep}{1mm}
 \begin{tabular}{c|c|c|c|c|c}
 \hline
 \textbf{Dataset} 	  & \textbf{\#Users}   & \textbf{\#Items} & \textbf{\#Instances} & \textbf{\#Features} & \textbf{\#Fields}  \\
 \hline
 Alipay   & 438,380 & 800,496 & 4,822,180 & 1,248,930 & 5  \\
 Tmall   & 415,800 & 565,888 & 4,573,800 & 994,771 & 8   \\
 Alimama  & 43,047 & 47,240 & 473,517 & 158,338 & 12 \\ 
 \hline
\end{tabular}
\label{tab:dataset}
	\vspace{-0.5cm}
\end{table}
\subsubsection{Baseline Models}
To verify the effectiveness of our proposed DG-ENN framework, we compare it with three groups of CTR prediction models: (A) feature interaction based models (LR \cite{lee2012estimating}, FM \cite{rendle2010factorization}, DeepFM \cite{guo2017deepfm}, PNN \cite{qu2018product}, AutoINT+ \cite{song2019autoint}); (B) user interest mining based models (DIN \cite{zhou2018deep}, DIEN \cite{zhou2019deep}); (C) GNN based models (GIN \cite{li2019graph}, FiGNN \cite{li2019fi}).

\subsubsection{Evaluation Metrics}
We adopt two widely-used evaluation metrics, namely \textit{AUC} and \textit{Logloss}~\cite{guo2017deepfm}, to evaluate the performance.
\textit{AUC} ($\uparrow$)  measures the goodness of assigning positive samples higher scores than randomly chosen negative samples.
A higher AUC value indicates a better performance.
\textit{Logloss} ($\downarrow$) measures the distance between the predicted scores and the true labels.
A lower Logloss value means a better model performance.

\subsubsection{Parameter Settings}
For fair comparison, we set embedding dimension of all models as 10, and batch size as 2000. We tune learning rate from \{1e-1,1e-2,1e-3,1e-4\}, $L_{2}$ from \{0,1e-1,1e-2,1e-3,1e-4,1e-5\}, and dropout ratio from 0 to 0.9. The deep layers for all models are \{400,400,400,1\}. The models are optimized with Adam optimizer \cite{kingma2014adam}. 
In addition to the above hyper-parameters for all models, we tune the GCN layer size for graph models in the range of \{1,2,3,4\}.
We use the validation set for tuning hyper-parameters, and the performance comparison is conducted on the testing set.
We run each experiments 5 times and report the average results.

\begin{table}[t]
\setlength{\abovecaptionskip}{0.1cm}
\setlength{\belowcaptionskip}{-0.0cm}
\centering
\caption{The overall comparison. 
$\star$ indicates a statistically significant level $p$-value<0.05 comparing DG-ENN with the best baseline (indicated by underlined numbers).}
\setlength{\abovecaptionskip}{0.2cm}
\setlength{\belowcaptionskip}{-0.0cm}
\setlength{\tabcolsep}{1mm}{
\small
\begin{tabular}{c|c|c|c|c|c|c}
\midrule[0.25ex]
Dataset &
\multicolumn{2}{c|}{Alipay} & 
\multicolumn{2}{c|}{Tmall} &
\multicolumn{2}{c}{Alimama} \\ \hline 
Model & AUC & Logloss &  AUC & Logloss  & AUC & Logloss \\\hline \hline
LR & 0.8196 & 0.2276 & 0.8760 & 0.1991 & 0.7207 & 0.2693  \\
FM  & 0.8498 & 0.2175 & 0.9026 & 0.1831 & 0.7396 & 0.2668  \\\hline
AutoInt+ & 0.8631 & 0.2147 & 0.9181 & 0.1730 & 0.7499 & 0.2611  \\
DeepFM  & 0.8648 & 0.2084 & 0.9155 & 0.1774 & 0.7653 & 0.2581  \\
PNN & \underline{0.8756} & \underline{0.2020} & \underline{0.9261} & \underline{0.1650} & \underline{0.7758} & \underline{0.2534}  \\ \hline
DIN & 0.8649 & 0.2081 & 0.9169 & 0.1761 & 0.7644 & 0.2584  \\
DIEN & 0.8731 & 0.2037 & 0.9235 & 0.1684 & 0.7710 & 0.2554  \\
 \hline
 GIN & 0.8645 & 0.2093 &0.9194 & 0.1716 & 0.7621 & 0.2595     \\
FiGNN & 0.8632 & 0.2121 & 0.9180 & 0.1753 & 0.7438 & 0.2635  \\ \hline
DG-ENN & $\textbf{0.9216}^{\star}$ & $\textbf{0.1674}^{\star}$ & $\textbf{0.9501}^{\star}$ & $\textbf{0.1399}^{\star}$ & $\textbf{0.8443}^{\star}$ & $\textbf{0.2254}^{\star}$ \\
\hline \hline
\end{tabular}}
\label{tab:ctraccuracy}
\end{table}

\subsection{Performance Comparison}\label{PerformanceComparison}
In this section, we compare the performance of DG-ENN with the state-of-the-art CTR prediction models.
Table \ref{tab:ctraccuracy} shows the experimental results of all compared models on three datasets.
We conduct Wilcoxon signed rank tests \cite{significant-test} to evaluate the statistical significance of DG-ENN with the best baseline algorithm. 
We have the following observations:
\begin{itemize}
    \item DG-ENN consistently yields the best performance for all datasets. 
    More precisely, DG-ENN outperforms the strongest baselines by \textbf{5.25\%}, \textbf{2.59\%} and \textbf{8.83\%} in terms of \textit{AUC} (\textit{17.13\%}, \textit{15.21\%} and \textit{11.05\%} in terms of \textit{Logloss}) on Alipay, Tmall and Alimama, respectively. 
    Possible reasons for the great improvement of DG-ENN over state-of-the-art CTR models may be the field-wise information propagation with attribute graph for alleviating the feature sparsity problem and the organized learning with user-item collaborative graph for behavior expanding across users and items.
    In contrast, most existing CTR methods ignore the rich relations existed in the data.
    We will further validate this observation in later experiments.
    \item LR performs worst among all baselines, which indicates that shallow linear combination of features is insufficient for CTR prediction. 
    FM performs better than LR, proves that the effectiveness of second-order feature interactions. 
    AutoInt+, DeepFM and PNN outperform FM, indicates that the modeling of high-order feature interactions is efficient for improving the performance of CTR prediction.
    DIN and DIEN achieve a comparable performance with DeepFM and PNN, demonstrates that user interest mining is also useful for representation learning. 
    \item GIN applies graph convolution on item-item graph to enrich user behaviors.
    However, it ignores the rich attribute information and the complex relations between users and items, thus it behaves much worse than DG-ENN.
    FiGNN employs graph convolution on field graph to model feature interactions.
    As no other relation information are introduced, it behaves no better than existing feature interaction based models. 
\end{itemize}


\subsection{Ablation Study of DG-ENN}
\label{componentComparison}
\begin{table}
\setlength{\abovecaptionskip}{0.2cm}
\setlength{\belowcaptionskip}{-0.0cm}
\caption{Compatibility of embedding enhancement.}
    \centering
\resizebox{\linewidth}{!}{
\small
\begin{tabular}{c|c|c|c|c|c|c}
\midrule[0.25ex]
Dataset &
\multicolumn{2}{c|}{Alipay} & 
\multicolumn{2}{c|}{Tmall} &
\multicolumn{2}{c}{Alimama} \\ \hline 
Model & AUC & Logloss &  AUC & Logloss  & AUC & Logloss \\\hline \hline
PNN & 0.8756 & 0.2020 & 0.9261 & 0.1650 & 0.7758 & 0.2534  \\
DG-PNN  & \textbf{0.9216} & \textbf{0.1674} & \textbf{0.9501} & \textbf{0.1399} & \textbf{0.8443} & \textbf{0.2254}  \\ \hline
DIN  & 0.8649 & 0.2081 & 0.9169 & 0.1761 & 0.7644 & 0.2584  \\
DG-DIN  & \textbf{0.9283} & \textbf{0.1608} & \textbf{0.9644} & \textbf{0.1176} & \textbf{0.8331} & \textbf{0.2317}  \\\hline
FiGNN & 0.8632 & 0.2121 & 0.9180 & 0.1753 & 0.7438 & 0.2635  \\
DG-FiGNN & \textbf{0.9115} & \textbf{0.1767} & \textbf{0.9432} & \textbf{0.1501} & \textbf{0.8155} & \textbf{0.2406}  \\
\hline \hline
\end{tabular}}
\label{tab:Effect of embedding enhancement}
\end{table}

\begin{table}
\setlength{\abovecaptionskip}{0.2cm}
\setlength{\belowcaptionskip}{-0.0cm}
\caption{Superiority of dual graph convolution.}
    \centering
\resizebox{\linewidth}{!}{
\small
\begin{tabular}{c|c|c|c|c|c|c}
\midrule[0.25ex]
Dataset &
\multicolumn{2}{c|}{Alipay} & 
\multicolumn{2}{c|}{Tmall} &
\multicolumn{2}{c}{Alimama} \\ \hline 
Model & AUC & Logloss &  AUC & Logloss  & AUC & Logloss \\\hline \hline
PNN & 0.8756 & 0.2020 & 0.9261 & 0.1650 & 0.7758 & 0.2534  \\\hline
GCN-PNN  & 0.9036 & 0.1842 & 0.9402 & 0.1542 & 0.7953 & 0.2487  \\
KGAT-PNN & 0.9096  & 0.1796 & 0.9426 & 0.1510 & 0.7968 & 0.2467   \\
HGAT-PNN  & \underline{0.9119} & \underline{0.1764} & \underline{0.9433} & \underline{0.1495} & \underline{0.8002} & \underline{0.2454}  \\\hline
DG-PNN  & \textbf{0.9216} & \textbf{0.1674} & \textbf{0.9501} & \textbf{0.1399} & \textbf{0.8443} & \textbf{0.2254}  \\ 
\hline\hline
\end{tabular}}
\label{tab:Superiority of dual graph convolution}
\end{table}

\begin{table}
\setlength{\abovecaptionskip}{0.2cm}
\setlength{\belowcaptionskip}{-0.0cm}
\caption{Effect of dual graph construction.}
    \centering
\resizebox{\linewidth}{!}{
\small
\begin{tabular}{c|c|c|c|c|c|c}
\midrule[0.25ex]
Dataset &
\multicolumn{2}{c|}{Alipay} & 
\multicolumn{2}{c|}{Tmall} &
\multicolumn{2}{c}{Alimama} \\ \hline 
Model & AUC & Logloss &  AUC & Logloss  & AUC & Logloss \\\hline \hline
PNN & 0.8756 & 0.2020 & 0.9261 & 0.1650 & 0.7758 & 0.2534  \\\hline
attribute graph & 0.9037 & 0.1831 & 0.9365 & 0.1545 & 0.8097 & 0.2428  \\
uu \& vv graph  & \underline{0.9122} & \underline{0.1753} & \underline{0.9438} & \underline{0.1473} & \underline{0.8232} & \underline{0.2353}  \\
uv graph & 0.9109 & 0.1771 & 0.9437 & 0.1477 & 0.8221 & 0.2371  \\\hline
DG-ENN & \textbf{0.9216} & \textbf{0.1674} & \textbf{0.9501} & \textbf{0.1399} & \textbf{0.8443} & \textbf{0.2254}  \\
\hline \hline
\end{tabular}}
\label{tab:Effect of dual graph construction}
\end{table}
In this section, we conduct a series of experiments to better understand the design rationality of our proposed DG-ENN.
\subsubsection{On the compatibility of embedding enhancement.}
To investigate the compatibility of our proposed dual graph enhanced embedding, we integrate PNN, DIN and FiGNN with the dual graph enhanced embedding, which we named as DG-PNN, DG-DIN and DG-FiGNN.
The experimental results are presented in Table \ref{tab:Effect of embedding enhancement}.
From these results, we can see that DG-PNN, DG-DIN and DG-FiGNN significantly outperform the original PNN, DIN and FiGNN models.
It validates the compatibility of our embedding enhancement approach by demonstrating its effectiveness on working with various popular CTR models.
This enhanced embedding is more informative with richer field-wise information and expanded user behaviors.

\subsubsection{On the superiority of dual graph convolution.}
To demonstrate the superiority of our proposed dual graph convolution module, we consider the variants of DG-PNN with different graph convolution models on our constructed graphs.
Specially, we compare dual graph convolution with GCN \cite{kipf2016semi}, KGAT \cite{wang2019kgat} and HGAT \cite{linmei2019heterogeneous}.
Noticed that the original GCN, KGAT and HGAT are not designed for CTR prediction.
We remove the prediction layer of these models and then apply them on our constructed graphs for embedding enhancement. We named these variants as GCN-PNN, KGAT-PNN and HGAT-PNN.
Table \ref{tab:Superiority of dual graph convolution} summarizes the results, from which we have the following findings:
\begin{itemize}
\item All these embedding enhanced models outperform the original PNN model, further verifies the effectiveness of embedding enhancement with relational information represented as graph.
\item KGAT-PNN behaves better than GCN-PNN on all three datasets. A possible reason is that GCN models the constructed graphs as a homogeneous graph, which ignores the different chasracteristics of differessnt fields while KGAT considers such differences.
\item HGAT-PNN outperforms KGAT-PNN on all three datasets.
This is because that HGAT-PNN utilizes all the graphs and models them in an heterogeneous manner, while KGAT only considers the collaborative graph and item-attribute graph.
\item DG-PNN consistently outperforms all baselines, which validates the superiority of our proposed dual graph convolution.
\end{itemize}

\subsubsection{On the effect of dual graph construction.}
We conduct experiments on three datasets to validate the effectiveness of the construction of attribute graph and collaborative graph.
We divide the collaborative graph into two parts: (1) user-user edges combined with item-item edges and (2) user-item edges, for detailed comparison.
Specially, we design four comparing variants: (1) DG-ENN only with the attribute graph (named attribute graph), (2) DG-ENN only with the user-user edges and item-item edges in the collaborative graph (named uu $\&$ vv graph), (3) DG-ENN only with the user-item edges in the collaborative graph (named uv graph) and (4) DG-ENN with neither attribute graph nor collaborative graph (that is PNN).
Table 
\ref{tab:Effect of dual graph construction} shows the comparison between different variants. 
We observe that PNN performs the worst in all these models, which proves the effectiveness of attribute graph and collaborative graph.
Moreover, we find that DG-ENN performs better than all the other models. 
It indicates that these attribute graph and collaborative graph are complementary to each other and can be combined together to improve the embedding quality and therefore boost the model performance.

\subsection{In-depth Analysis on Graph Modeling}\label{componentComparison}

\subsubsection{Impact of Aggregators.}
\begin{figure}[htp]
	\centering
	\setlength{\belowcaptionskip}{-0.3cm}
	\setlength{\abovecaptionskip}{0cm}
	\includegraphics[width=0.45\textwidth]{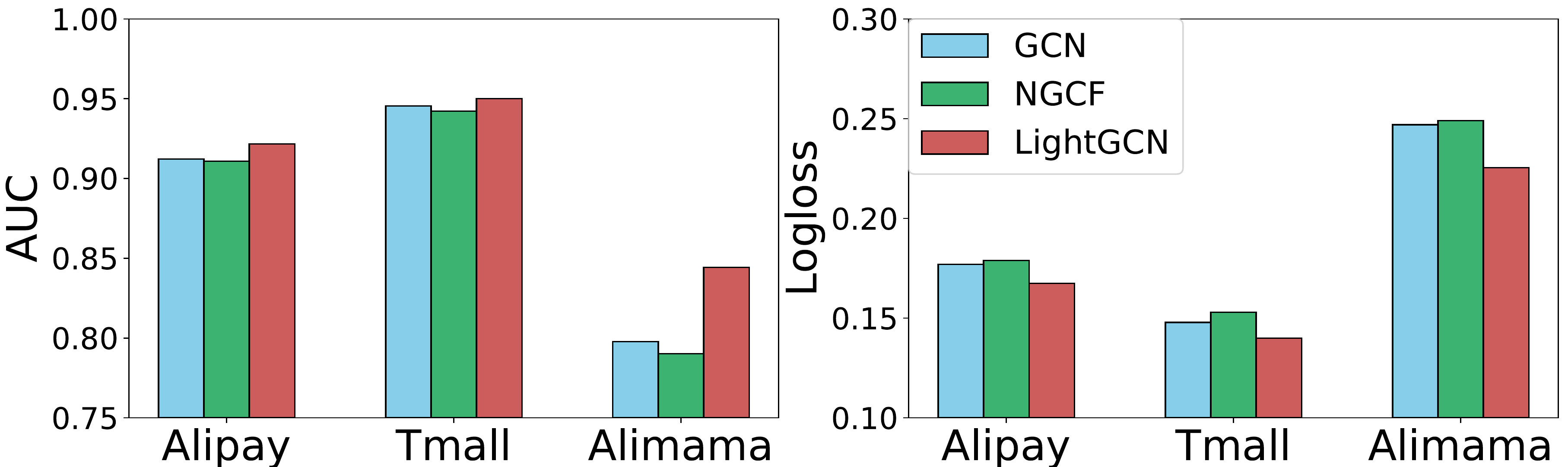}
	\caption{Impact of Aggregators.}
	\label{fig:figureaggregators}
\end{figure}

To explore the impact of different aggregators, as formulated in Equation \ref{GCN}-\ref{lightgcn}, we compare the performance of our proposed model with different aggregators.
Figure \ref{fig:figureaggregators} summarizes the experimental results.
We can see that GCN aggregator performs better than NGCF aggregator on all datasets.
A possible reason is that additional feature interactions between central node and neighbor nodes introduced by NGCF aggregator makes it easy to overfit.
Moreover, we can see that LightGCN aggregator which removes both the weight matrix and activation function achieves the best performance on all datasets.

\subsubsection{Impacts of Attribute Information Exploitation}
\begin{figure}[htp]
	\centering
	\setlength{\belowcaptionskip}{-0.3cm}
	\setlength{\abovecaptionskip}{0cm}
	\includegraphics[width=0.45\textwidth]{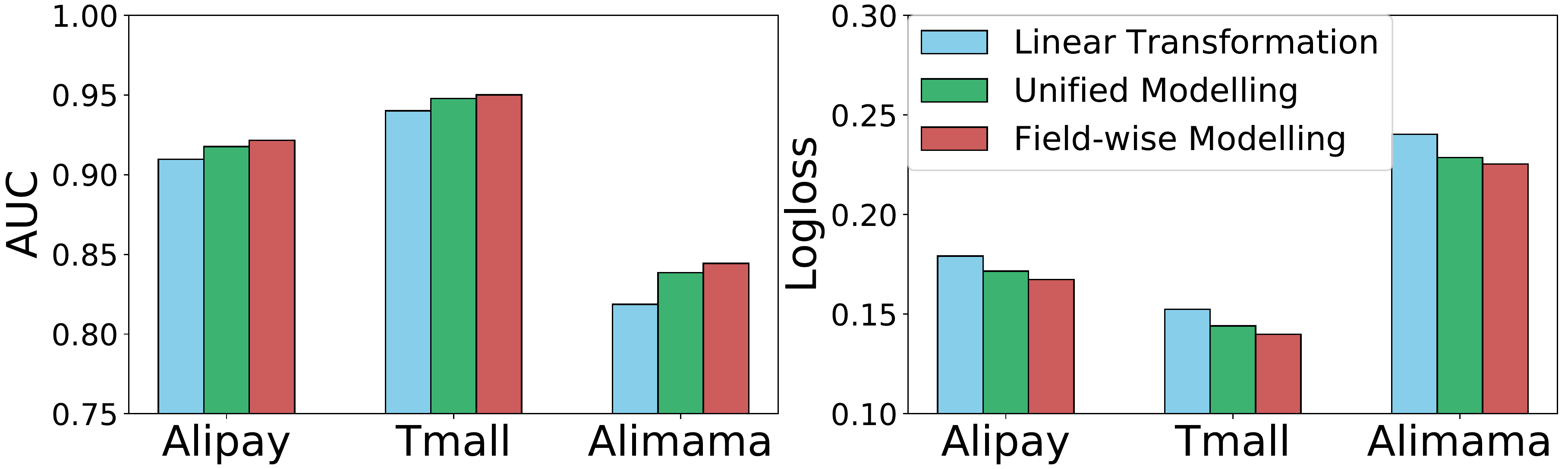}
	\caption{Impact of Attribute Information Exploitation.}
	\label{fig:figureInformationExploitation}
\end{figure}

To verify the effectiveness of our divide-and-conquer strategy to integrate different attribute information, as explained in Section \ref{Attribute Graph Convolution}, 
we replace our proposed attribute graph convolution module with other two alternatives: (1) using the linear transformation of ID embedding and attribute embeddings as the refined user/item representation \cite{kipf2016semi,berg2017graph}, (2) modeling the different fields of attributes without considering their fields.
Figure \ref{fig:figureInformationExploitation} shows the experimental results, we can see that the first alternative gets the worst performance, proving the effectiveness of modeling attributes as graphs.
Besides, modeling the different fields of attributes without considering their fields (i.e., the second alternative) performs worse than our model, verifies the necessary of modeling field-wise information individually.

\subsubsection{Impacts of Collaborative Signal Exploiting}
\begin{figure}[htp]
	\centering
	\setlength{\belowcaptionskip}{-0.3cm}
	\setlength{\abovecaptionskip}{0cm}
	\includegraphics[width=0.45\textwidth]{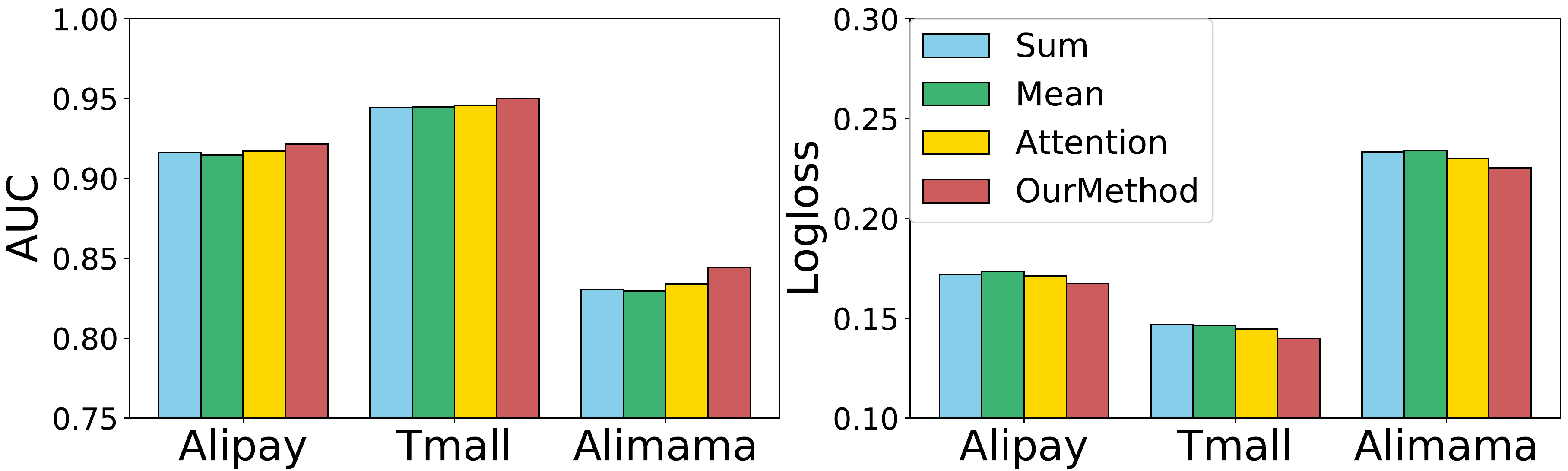}
	\caption{Impacts of Collaborative Signal Exploiting.}
	\label{fig:figureCollaborativeSignalExploiting}
\end{figure}

To validate the superiority of our design of organized learning for the collaborative graph, as explained in Section \ref{Collaborative Graph Convolution}.
We conduct three different operations on the aggregated embeddings from multiple types of edges: 
(1) element-sum operation; (2) element-mean operation; (3)  attention operation.
From the results in Figure \ref{fig:figureCollaborativeSignalExploiting}, we can see that our DG-ENN obtains the best results.
Besides, we find that attention operation achieves the second best results.
\subsection{Case Study}
In this part, we conduct experiments to verify that our model can solve the problem of feature sparsity and behavior sparsity.

\subsubsection{Feature Sparsity Analysis}
In order to prove that our model can solve the feature sparsity well, we select instances in the test set containing one of the four features with low frequency in the training set. The four chosen features are presented in Table \ref{tab:CaseStudy}, where they are represented by feature fields with subscripts of desensitization information. We report the performance (i.e., Logloss) of PNN and DG-PNN on the selected test instances in Table \ref{tab:CaseStudy}. We can find that DG-PNN achieves significant performance improvement on the test samples with sparse features, compared to PNN. This result demonstrates that our proposed  dual graph enhanced embedding alleviates the feature sparsity issue.


\begin{table}[htp]
\setlength{\abovecaptionskip}{0.1cm}
\setlength{\belowcaptionskip}{-0.0cm}
\caption{Feature Sparsity Analysis in Alimama.}
    \centering
\scalebox{0.9}
{
\small
\begin{tabular}{c|c|c|c}
\midrule[0.25ex]
Feature &  Frequency & PNN (Logloss) & DG-PNN (Logloss) \\ \hline 
Brand\_{1} & 12 & 0.3502 & 0.2868 \\
Brand\_{2} & 5 & 0.3218 & 0.3111  \\ 
Cate\_{1} & 8 & 0.6125 & 0.5645    \\ 
Cate\_{2} & 9 & 0.0851 & 0.0223    \\ 
\hline \hline
\end{tabular}}
\label{tab:CaseStudy}
\end{table}
\subsubsection{Behavior Sparsity Analysis.}
\begin{figure}
	\centering
	\setlength{\belowcaptionskip}{-0.3cm}
	\setlength{\abovecaptionskip}{0cm}
	\includegraphics[width=0.4\textwidth]{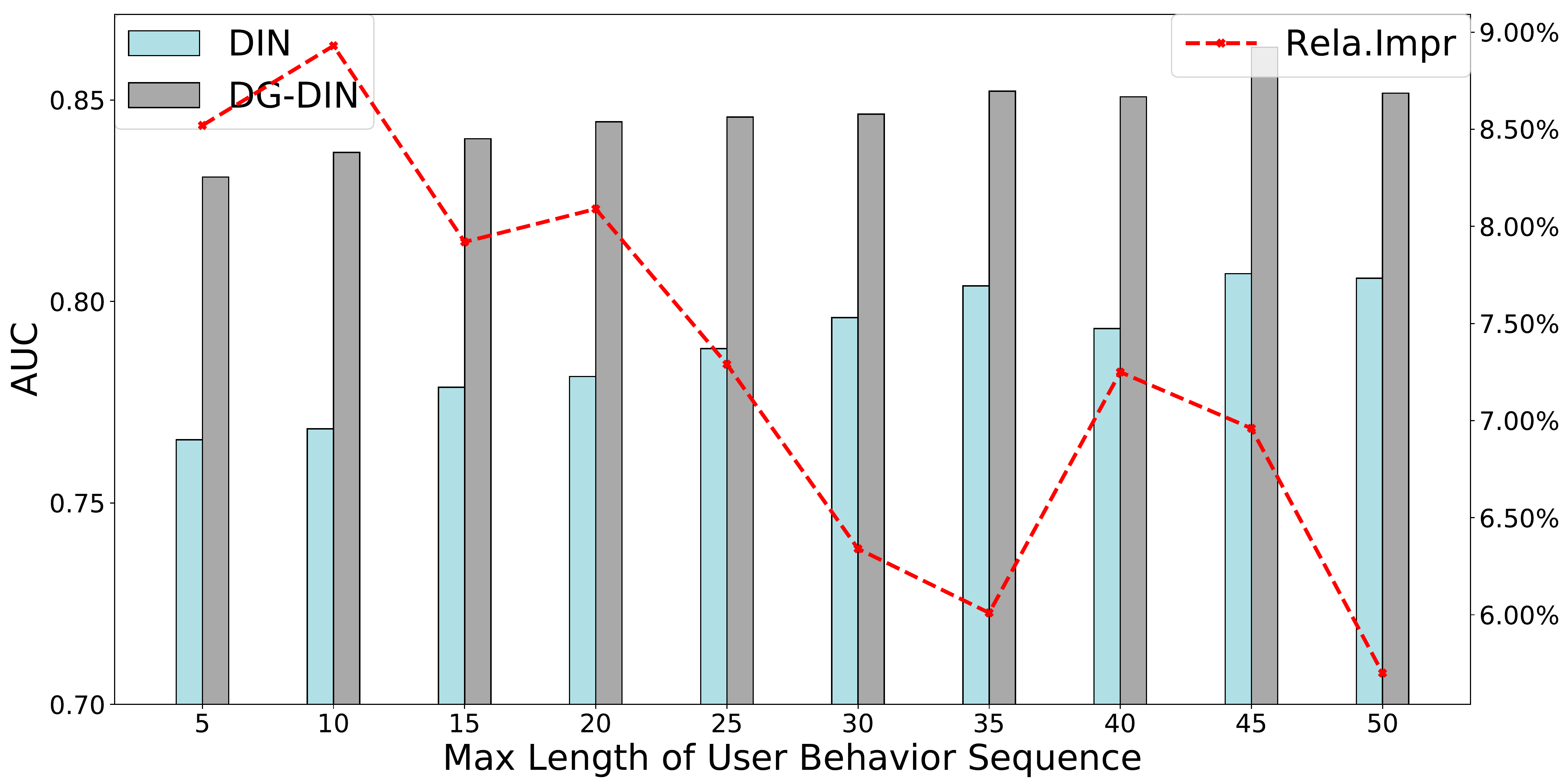}
	\caption{Behavior Sparsity Analysis.}
	\label{fig:BehaviorSparsityAnalysis}
\end{figure}

Besides, the behavior sparsity problem can also be solved well by our model. We choose Alipay dataset for experiment because this dataset includes  less attribute information which may make noise for behavior sparsity analysis. Figure \ref{fig:BehaviorSparsityAnalysis} shows the performance comparison between DIN and DG-DIN with respect to different lengths of user behavior sequences. 
The result shows that the relative improvement of DG-DIN over DIN is more significant when length of user behavior sequence is less. That is to say, our proposed dual graph enhanced embedding alleviates the behavior sparsity issue.

\section{Conclusions}
In this paper, we focus on exploiting the graph representation learning to alleviate the feature sparsity and behavior sparsity problems for existing CTR models.
We propose a novel dual graph enhanced neural network based on attribute graph and collaborative graph. 
On the one hand, to learn the feature representation from attribute graph effectively, we propose a divide-and-conquer learning strategy to perform field-wise attribute modeling.
On the other hand, to model the complex user-item relation for behavior expanding, we design a organized learning strategy inspired by curriculum-learning to learn the correlations within users/items and also between users and items. 
The extensive experiments on three real-world datasets have demonstrated the superiority of our proposed DG-ENN over the state-of-the-art methods. Moreover, the proposed dual graph enhanced embedding is able to work collaboratively with various deep CTR models to boost their performance.



\balance


\bibliographystyle{ACM-Reference-Format}
\bibliography{sample-bibliography}

\end{document}